\documentstyle{article}\begin{document}

\centerline{\bf  A note on the Exact Green function for a quantum system}
\centerline{ \bf decorated by two or more impurities.}

\vskip .5in

\centerline{M.L. Glasser}

\vskip .3in

\centerline{ Dpto de F\'isica Te\'orica, Facultad de Ciencias}
\centerline{Universidad de Valladolid}\vskip .2in

\centerline{Physics Department, Clarkson University,}
\centerline{Potsdam, NY  13699-5820} (USA)
\vskip 1in
\centerline{\bf Abstract}\vskip .1in
\begin{quote}
 The exact Green function is constructed for a quantum system, with known Green function, which is decorated by two  delta function impurities.It is shown that when two such impurities coincide they behave as a single singular potential with combined amplitude. The results are extended to N impurities.
 
 \end{quote}
 

 \newpage
 \section{Introduction}
 The one dimensional harmonic oscillator or square well, for example,  for which the energy -dependent Green function $G_0(x,x';E)$ is known,  have been taken for many years as solvable models for semi-conductor quantum wells[1]. Frequently delta function potentials are placed at various points to simulate defects or impurities. In the case of a single such potential, $V(x)=\lambda \delta(x-a)$, the Green function for the composite system is known to be[2]
 $$ G(x,x';E)=G_0(x,x')+\lambda\frac{G_0(x,a;E)G_0(a,x')}{1-\lambda G_0(a,a;E)}\eqno(1)$$
 
 In this note a corresponding formula is derived for the case $V(x)=\lambda\delta(x-a)+\mu\delta(x-b).$
 
 \section{Calculation}
 We first note that the same argument can be used for the time dependent-, as well as the energy-dependent Green functions, so we shall omit the third argument and write simply $G(x,x')$.
 
 Beginning with the Dyson equation, noting that $G_0(x,y)=G_0(y,x)$
 $$G(x,x')=G_0(x,x')+\int G_0(x,y)V(y)G(y,x')dy,\eqno(2)$$
 where the integration extends over the system domain, one has the set of equations
 $$G(x,x')=G_0(x,x')+\lambda G_0(x,a)G(a,x')+\mu G_0(x,b)G(b,x')\eqno(3)$$
 $$G(a,x')=G_0(a,x')+\lambda G_0(a,a)G(a,x')+\mu G_0(a,b)G(b,x'),\eqno(4)$$
 $$G(b,x')=G_0(b,x')+\lambda G_0(a,b)G(a,x')+\mu G_0(b,b)G(b,x').\eqno(5)$$
 The linear equations (4) annd (5) are  easily solved for $G(a,x')$ and $G(b,x'):$
 $$G(a,x')=\frac{G_0(a,x')+\mu[G_0(b,x')G_0(a,b)-G_0(a,x')G_0(b,b)]}{D}\eqno(6)$$
 $$G(b,x')=\frac{G_0(b,x')+\lambda[G_0(a,x')G_0(a,b)-G_0(b,x')G_0(a,a)]}{D}\eqno(7)$$
 with
 $$D=[1-\lambda G_0(a,a)][1-\mu G_0(b,b)]-\lambda\mu [G_0(a,b)]^2.\eqno(8)$$
 By inserting (6) and (7) into (3) we obtain the desired expression
 \newpage
 $$G(x,x')=G_0(x,x')$$
 $$+\frac{1}{D}\{\lambda G_0(x,a)G_0(a,x')+\mu G_0(x,b)G_0(b,x')$$
 $$+\lambda\mu[G_0(x,a)\left(G_0(a,b)G_0(a,x')-G_0(b,b)G_0(b,x')\right)$$
 $$+G_0(x,b)\left(G_0(a,b)G_0(a,x')-G_0(a,a)G_0(a,x')\right)]\}.\eqno(9)$$
 
 \section{Discussion}
 By setting $\lambda$ to $0$  (9) reduces to (1), proving this expression as well. The most salient feature of (9) is the denominator $D$ whose zeros form the exact spectrum of the composite system.  For example, when $a$ and $b$ coincide, $D$ reduces to $1-(\lambda+\mu)G(a,a)$ and (9) reduces to (1) with $\lambda$ replaced by the  amplitude $\lambda+\mu$. I.e. the two impurities combine to form one with combined amplitude. This generalizes  the result of Rinaldi and  Fasssari[3],  for two identical defects symmetrically placed with respect to the center of a harmonic oscillator.
 
    Two further points can be made. Nothing in the derivation of (9) restricts it to the line. If we accept the standard definition
    $\delta(\vec{x})=\Pi_{j=1}^d\delta(x_j)$, then (9), and its consequences,  are  valid for $d$-dimensional  quantum systems. This has been proven function-theoretically for the three dimensional harmonic oscillator with two symmetrically placed identical impurities by Albeverio, Fassari and Rinaldi4].
    
    A second observation is that $D$ is simply the Cramer determinant for the pair of simultaneous linear equations (4) and (5). In the case of  impurity potential
    $$V(x)=\sum_{j=1}^N \lambda_j\delta(x-a_j)\eqno(10)$$ 
    there will be $N$ such equations and the determinant is easily evaluated. Thus, for $N=3$
    
    $$D=\prod_{j=1}^3[1-\lambda_jG_0(a_j,a_j)]-\sum_{i< j}\lambda_i\lambda_j G_0(a_i,a_j)G_0(a_j,a_i)$$
    $$+\lambda_1\lambda_2\lambda_3 G_0(a_1,a_2)G(a_2,a_3)G_0(a_3,a_1).\eqno(11):$$
    This reduces to the $N=1$ and $N=2$ cases appropriately and shows that any two coinciding impurities coalesce as indicated above.
    
    Finally, by letting N become infinite, the analogue of (11) might offer a new approach to  Kronig-Penney-type systems for periodic or random unit cells.
\vskip .4in    
    \noindent
   {\bf Acknowledgement}:  The author thanks Prof. S. Fassari and Prof. L.M. Nieto for helpful comments and  acknowledges  the financial support of MINECO (Project MTM2014-57129-C2-1-P) 
and Junta de Castilla y  Leon (VA057U16)).
\vskip .8in

\noindent
 {\bf References}\vskip .1in

\noindent
[1] S. Albeverio, F. Gesztesy, R.Hoegh-Krohn and H. Holden, {\it Solvabe Models in Quantum Mechanics}, [AMS-Chelsea,  Providence (2004)]

\vskip .1in\noindent
[2] M.L. Glasser and L.M. Nieto, {\it The energy level structures of a variety of one-dimensional confining potentials and the effects of a local singular perturbation}, Can. J. Physics {\bf 93}, 1-9 (2015).
\vskip ,1in

\noindent
[3] 3] S. Fassari and F. Rinaldi, {\it On the spectrum of the Schroedinger Hamiltonian of the one-dimensional harmonic oscillator perturbed by two identical attractive point interactions}, Reports on Mathematical Physics {\bf 69}, 353-370 (2012).\vskip .1in

\noindent
[4] S.Albeverio, S. Fassari and F. Rinaldi,{\it Spectral properties of a symmetric three-dimensional
quantum dot with a pair of identical attaractive impurities
symmetrically situated around the origin}, Nanosystems, Physics, Chemistry and Mathematics {\bf 7},268-289 (2016).

 \end{document}